\documentclass[a4paper]{article}

\usepackage[english]{babel}
\usepackage[utf8]{inputenc}
\usepackage{amsmath}
\usepackage{setspace}
\usepackage{setspace}
\usepackage{graphicx}
\usepackage{caption}
\usepackage[margin=0.5in]{geometry}

\title{On Mortgages and Refinancing}

\author{Khizar Qureshi \thanks{In collaboration with Morgan Stanley Quantitative Finance, Strats}, Cheng Su}

\date{\today}

\begin{document}
\maketitle
\doublespacing

\begin{abstract}

In general, homeowners refinance in response to a decrease in interest rates, as their borrowing costs are lowered. However, it is worth investigating the effects of refinancing after taking the underlying costs into consideration. Here we develop a synthetic mortgage calculator that sufficiently accounts for such costs and the implications on new monthly payments. To confirm the accuracy of the calculator, we simulate the effects of refinancing over 15 and 30 year periods. We then model the effects of refinancing as risk to the issuer of the mortgage, as there is negative duration associated with shifts in the interest rate. Furthermore, we investigate the effects on the swap market as well as the treasury bond market. We model stochastic interest rates using the Vasicek model.

\end{abstract}
\doublespace
\section{Introduction}

The decision to refinance mortgages is often undermined, and unfortunately, misunderstood by many homeowners. In particular, one overlooked consideration is the closing costs associated with refinancing. Modern real estate finance has developed an intuition for the decision, and subjects it to several factors, including: magnitude of closing costs, change in mortgage rate, borrowing costs, and the remaining time of the mortgage. Interestingly enough, a decrease in mortgage rate or borrowing costs should not always be entertained with the decision to refinance. As we will see, there exists a threshold for the time remaining on the mortgage to validate the net present value of such a decision. 

First, we develop a synthetic mortgage calculator with two main features. The first feature is very simple, and conveys to the homeowner the monthly payments associated with a traditional mortgage. The second feature, however, provides the proportions of the mortgage that are allocated towards the interest and principal payments. Upon completion of this mortgage calculator, we will extend its features, and simulate scenarios of refinancing.

The remainder of the paper is organized as follows: an interpretation of the synthetic mortgage calculator includes the time-variant distribution of a monthly mortgage payment to its interest and principal components will be followed by the discrete potential monthly savings available to a homeowner through refinancing. Here we will depict the conditions under which an option to refinance should be exercised. Next, we will simulate the growth and breakeven in net present value of a repayment option. We will see that the initial costs are eventually overtaken by the savings in decreased monthly interest payments. The completion of the primal purpose will lead to a discussion of implications on contract duration, bond pricing, MBS hedging as well as short-term interest rates. 
\subsection{Synthetic Mortgage Calculator}

Here we present a synthetic mortgage calculator. Given a rate x, a term of N years, a notional of B dollars, we will determine the monthly payment. Further, we will determine the portion of the payment that goes towards paying the principal and the portion of the payment that goes towards paying the interest.

\begin{equation} 
B = \sum_{i=1}^{12N} \frac{M}{[1+ \frac{x}{12}]^i} 
\end{equation}

The effective annual interest rate is therefore:

\begin{equation} EAR = [1+\frac{x}{12}]^{12} -1 \end{equation}

From the monthly payment, M, the interest expense, $E_{i}$, is equivalent to:

\begin{equation} E_i = \frac{x}{12}  B_i \end{equation}

where $B_i$ is the outstanding balance. It follows that the remainder of the payment goes towards paying the principal, $E_p$:

\begin{equation} E_p = M - \frac{x B_i}{12}
\end{equation} 

The proportions, $X_i$ and $X_p$ are: $X_p = 1 - \frac{x  B_i}{12M}$ and $X_i = \frac{x B_i}{12M}$. \\

\begin{center}
\includegraphics[width=0.7\textwidth]{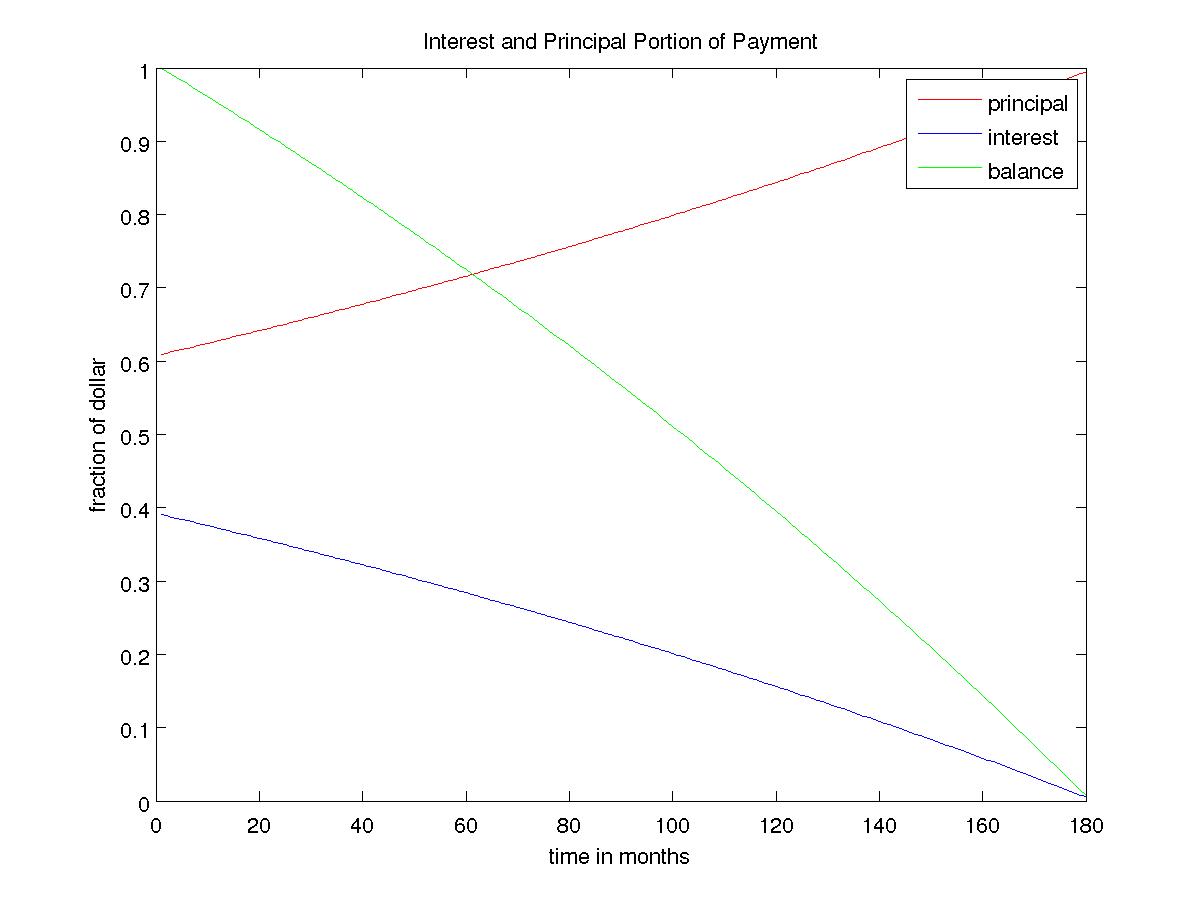}%
\captionof{figure}{The figure above illustrates the portions of the monthly payment that are paid towards interest and principal for a 15-year mortgage with a rate of 3.29 percent. Notice that the portion of the payment allocated towards principal is far greater than that allocated towards interest. }\label{labelname}%
\end{center}


\begin{center}
\includegraphics[width=0.7\textwidth]{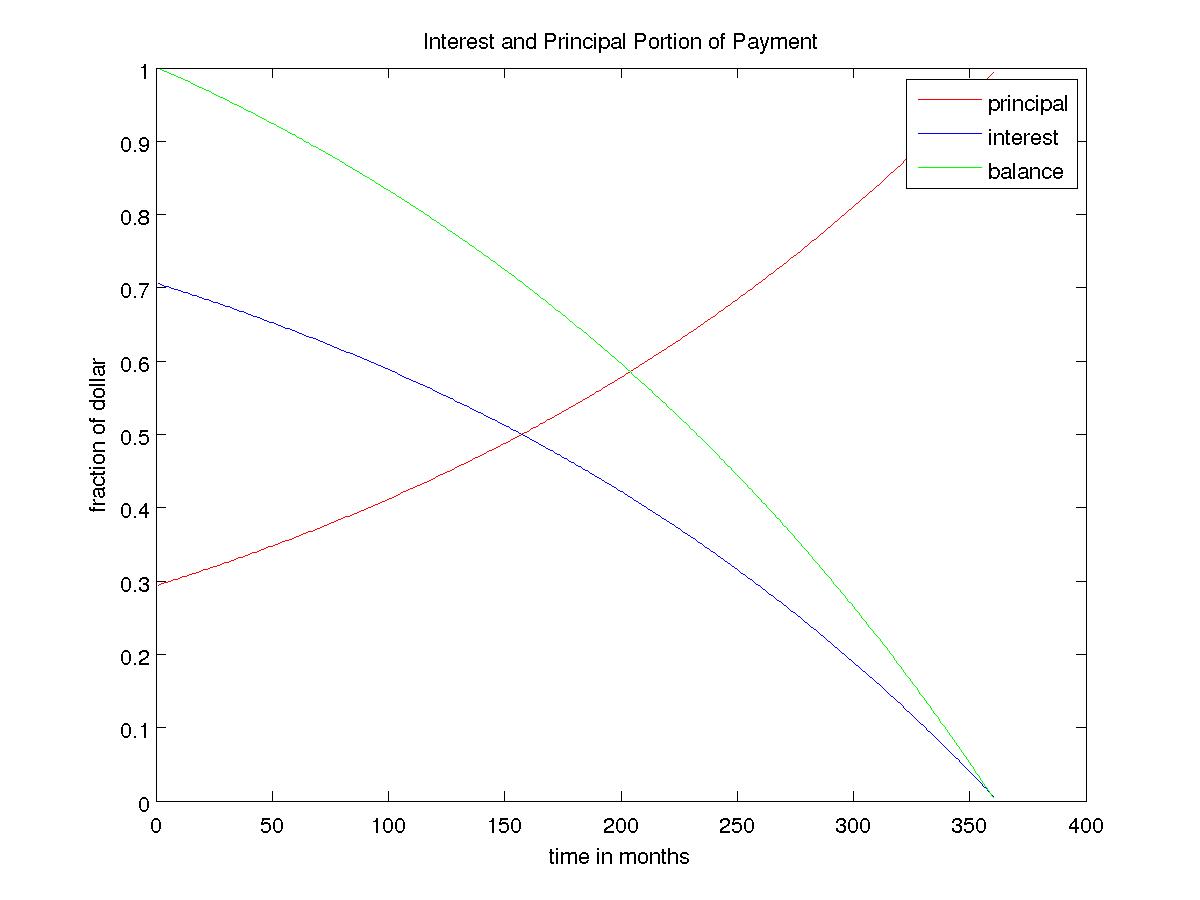}%
\captionof{figure}{The figure above illustrates the portions of the monthly payment that are paid towards interest and principal for a 30-year mortgage with a rate of 4.03 percent. Notice that initially, the portion of the payment allocated towards the interest is greater than that allocated towards the principal (unlike the 15-year mortgage). However, as time progresses, the payments are weighted towards the principal component.}\label{labelname}%
\end{center}

Note that the mortgage payment is constant on a monthly basis. Further, it is clear that the percent of principle payment increases over time, while the percent of interest payment decreases over time. This payment procedure is also known as an amortization schedule.

\section{Refinancing}

\subsection{Cash Flows with Prepayments}

Let S = the monthly prepayment rate for n months. Also suppose we have the following variables in month n without prepayment risk: Balance, $B_n$, Interest, $I_n$, and Principal, $P_n$. It follows that $Q_n = \prod_{i=1}^n (1-S_n)$. The total monthly payment in month n is simply:

\begin{equation} \hat{M_n} = \frac{\hat{B_n} i (1+i)^{N-n+1}}{(1+i)^{N-n+1} -1} = M_n Q_{n-1} \end{equation}  

with a scheduled principal portion of the monthly payment equivalent to:

\begin{equation} \hat{P_n} = \frac{\hat{B_n}i}{(1+i)^{N-n+1} -1} = P_n Q_{n-1} \end{equation}

and an interest portion equivalent to

\begin{equation} \hat{I_n} = \hat{B_n} i = I_n Q_{n-1} \end{equation}

and unscheduled principal payment in month n equivalent to:

\begin{equation} P^*_n = S(\hat{B_{n-1}} - \hat{P_n}) \end{equation}

and a remaining balance at the end of the month equivalent to:

\begin{equation}  \hat{B_n} = B_{n-1} -\hat{P_n} - P^*_n = B_n Q_n \end{equation}

Through refinancing, the interest savings over time is simply $\sum_{i=n}^N \hat{I_i} - I_i$. It follows that for longer maturities, the accrued interest savings on a mortgage are higher. 

\subsection{Savings}

We now consider the option of refinancing such a mortgage. Namely, given mortgages A and B, and a refinancing cost C, should mortgage A be refinanced to mortgage B for immediate cost C? And if it should, what is the time required to break even? We will see that such a decision is dependent on factors such as: current interest rate, new interest rate, closing costs, and the time planned to remain in the home.

Given a new interest rate, $X_{new}$, the monthly payment is:

\begin{equation} M_{new} = B \frac{x_{new}}{12}  \bigg[1- \frac{1}{[1+\frac{x_{new}}{12}]^{12N}} \bigg]\end{equation}

It follows that the monthly savings, $S_i$, is simply $\hat{I_i} - I_i$.

The mortgage should only be refinanced if the net present value of doing so is positive. That is:

\begin{equation} NPV = -C + \sum_{i=n}^N [\hat{I_n}-I_n]  \geq 0 \end{equation}

Therefore, the number of months required for the interest and principal mortgage insurance savings, S to exceed closing costs, C are:

\begin{equation} t_{Months} = \frac{C}{S_{i}} \end{equation}

However, there exist a few situations in which refinancing should not be done, even with lowered interest rates
\begin{itemize}
\item only a short term structure for the loan remains
\item prepayment penalties exceed expected savings
\item mortgage rates are expected to drop further
\end{itemize}

\subsection{Simulations}

To develop a practical understanding of refinancing, we simulate a specific example. Suppose there is a 20-year outstanding mortgage with an initial loan amount of 100,000. Further, assume that there is a state of nature in which there is a 1 percent drop in interest rates. The homeowner is tempted to borrow at this lower interest rate, and prepay, but should be wary of closing costs. In particular, the homeowner should ponder the following: Given the closing costs, and the remaining time period on the mortgage, is the net present value of refinancing positive? For any rational monetary decision, this is a strict condition. Below is a figure that illustrates the results of the simulation. Notice that due to closing costs, the initial $NPV_{t=0}$, is negative. However, 15 months into the refinancing decision, the NPV breaks even. 

\begin{center}
\includegraphics[width=1.0\textwidth]{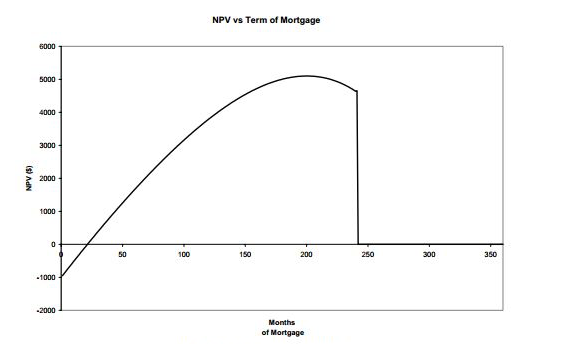}%
\captionof{figure}{The figure above illustrates the number of months [15] required for the net present value to breakeven during refinancing. Suppose there is a 20-year (outstanding) mortgage with an initial loan of 100,000. Although there is a 1 percent drop in interest rates, there is a remaining balance of 85812, and closing costs are approximately 1000. The new loan issued through refinancing is simply the sum of the remaining balance and the quoted closing costs.}\label{labelname}%
\end{center}

Consider the very same mortgage, but with an outstanding term of less than 15 months. It follows that with the same closing costs and base mortgage rate, the net present value of refinancing will surely be negative. That is, $\forall t<15years, NPV_{T} <0$. This is because the payment structure is not allotted sufficient time to break even the costs. In scenarios such as these, the mortgage should never be refinanced.

Next, we generalize our simulations to illustrate the difference in activity for a 15-year and 30-year mortgage. In particular, with and without closing costs, we try to determine the effects of refinancing on the payment structure for two traditional contract lengths. First, we consider an unrealistic case in which rates are lowered, and no closing fees are imposed:

\begin{center}
\includegraphics[width=0.7\textwidth]{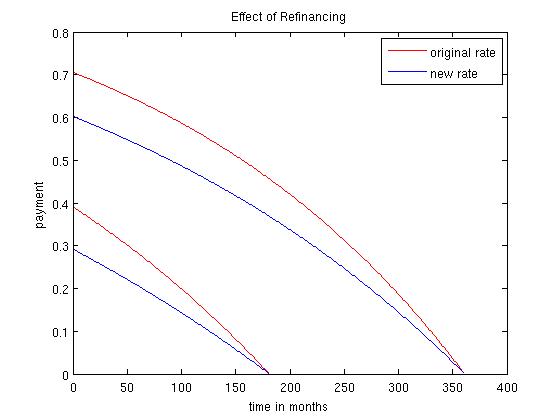}%
\captionof{figure}{The graph above illustrates the reduction in monthly coupon with refinancing. Here we assume zero costs to refinance and present the effects in both a 15-year and 30-year mortgage. Notice that because of the unrealistic zero costs, the effective monthly payment is reduced significantly. However, as time progresses, the monthly payment converges. This holds for both a 15-year and 30-year mortgage.}\label{labelname}%
\end{center}

\begin{center}
\includegraphics[width=0.7\textwidth]{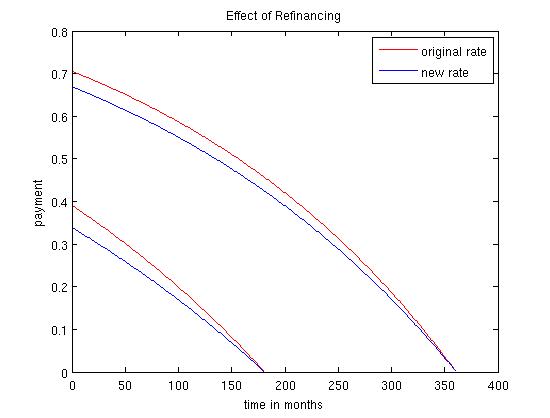}%
\captionof{figure}{The graph above illustrates the reduction in monthly coupon with refinancing. Here we assume fixed, non-zero costs to refinance and present the effects in both a 15-year and 30-year mortgage. In particular, to account for the costs, the 15-year and 30-year mortgage rates are given a multiplier $\lambda=1.20$. Effectively, this increases the capital commitment to replicate closing costs. Notice that due to the newly introduced costs, the effective difference in monthly payments has narrowed significantly. It is also observable that the difference narrows more slowly with the 30-year mortgage, than it does with the 15-year mortgage. }\label{labelname}%
\end{center}

The simulations above aid in the development of an intuition for mortgage activity for the following states:

\begin{itemize}
\item Refinancing without closing costs (15-year mortgage)
\item Refinancing without closing costs (30-year mortgage)
\item Refinancing with closing costs, $\lambda$ (15-year mortgage)
\item Refinancing with closing costs, $\lambda$ (30-year mortgage)
\end{itemize}

It is clear that closing costs, depending on magnitude may significantly reduce the perceived benefits of refinancing through means of increasing the effective monthly payments.

\section{Negative Duration Risk}

In most mortgages borrowers are given prepayment options, and are typically exercised when interest rates fall. Homeowners are able to borrow at these lower interest rates, and in most cases, effectively reduce their monthly payments. However, this results in a negative convexity for Mortgage-Backed Securities (MBS). As interest rates fall, the duration for Mortgage-Backed Securities actually shortens. Inversely, an increase in interest rates discourages homeowners from borrowing and refinancing, resulting in a positive duration. The consequences of such an effect are quite significant, and may be readily observed. The equation below models the effect on duration, D, with a change in yield (interest rates), y (Myers, 2011). 

\begin{equation} D = -\frac{dP}{dY} \frac{1+y}{P} \end{equation}

Notice that with an increase in interest rates, duration typically decreases. However, in the case of MBS, duration decreases due to prepayment and refinancing effects.

As a response to the shortening duration, market participants hedge with interest rate swaps, and opt to receive the fixed leg, or equivalently, terminate any floating receivables. Moreover, such an effect can result in buying pressure on the "swap bond", which in turn results in temporary decrease in the swap spread. The mechanism above is opposite of traditional bond activity, justifying the compensation market participants demand for negative duration risk. This level of intuition may be valuable when judging credit market activity.

It is critical to not only understand the effects of duration on the credit market, but also how interest rates vary over time. In particular, we explore the Vasicek model, which correctly hypothesizes the stochastic movement of interest rates. We then extend the model parameters to bond pricing, which is of interest to market participants who seek to proliferate from interest rate movement. Finally, we return to the idea of hedging, and why the MBS swap market is so liquid. 

\section{Vasicek Model for Interest Rates}
The Vasicek Model in a continuous-time term-structure model for short rates. It follows a stochastic mean reversion process as follows (Vasicek, 1977):

\begin{equation} dr_{t} = \kappa (\bar{r} - r_t)dr +\sigma dW_t \end{equation}

where $r_t$ is the short rate, $\kappa$ the mean reversion coefficient, and dW captures a Wiener Process \cite{Vasicek}. 

\subsection{Implication on Bond Pricing}

The price of a T-year Zero Coupon Bond may be determined by: 
\begin{equation}  P = e^{\alpha + r \beta} \end{equation}

where $\beta = \frac{e^{-\kappa T}-1}{\kappa}$ and $\alpha = \bar{r}\bigg[-\beta -T\bigg]+\frac{\sigma^2}{2 \kappa^2}\bigg[\frac{1-e^{-2 \kappa T}}{2 \kappa} + \frac{2 \beta}{\kappa} +T\bigg]$ 

Once $\kappa$ is determined from the Vasicek model, it may be used to determine the resulting bond price. It is readily observable that a larger value of $\kappa$, or equivalently, a faster mean reversion process implies lower bond prices. This increase in yield  translates to a narrowing in swap spreads. 

\subsection{Implication on Short Rate}
Consider a three-month T-bill with rate $r_t$ at time t. The vasicek model, in discrete form, implies that at  time t+1, the T-bill will carry rate:

\begin{equation} r_{t+1} =\kappa \bar{r} \Delta + (1- \kappa \Delta)r_{t} + \sigma \Delta^{1/2} \epsilon_{t+1} \end{equation} 
where $\sigma$ is the one-month volatility, $\Delta$ a change in time of $\frac{1}{12}$ years, and $\epsilon_{t+1}$ a small shock such that $\epsilon \in N(0,1)$. Note that in practice, the short rate would follow a multi-factor model.

\subsection{Implication on MBS Hedging }

Consider a swap spread $\lambda - T$ where $\lambda$ is the swap rate for a mortgage-backed security. It follows that with in increase in short rate rate, the swap spread would decrease. This is attributable to two factors. One straightforward reason is that the bond yield would increase in response to the increase in short rate. The second reason, however, is that with an increase in duration, market participants will choose to sell duration through means of paying the fixed leg of the swap (or equivalently, sell current receivables), which effectively lowers the swap rate. A decrease in the swap rate and an increase in bond yield result in drastic decreases in the swap spread. Inversely, decreases in interest rates result in large increases in the swap spread. 

\section{Conclusion}

A decrease in interest rates provides incentive to refinance mortgages. However, as we have seen, an important consideration is the savings on the mortgage taking into account the costs for refinancing. Only if the 15-year or 30-year sum of the change in monthly payments exceed the borrowing costs should the mortgage be refinanced. In particular:
\begin{itemize}
\item Differential changes in the interest rates should not be reasons to refinance.
\item There exists a minimum time $T>0$ for the remaining life of the mortgage only under which it is optimal to refinance
\item Closing costs are almost always non-zero, so the preceeding conditions are therefore $strict$
\end{itemize}

Refinancing and prepayment also provide sources of risk to mortgage issuers in the form of negative duration. Issuers face the risk of having the sum of the time-weighted payments reduced. Compensation to carry such risks are likely factored into the mortgage rate offered to the homeowner depending on his/her individual propensity to refinance or issue prepayments.\\

Changes in the interest rate have implications on the MBS market, treasury bond market, and swap market. A decrease in the interest rate can lower the prices of mortgage-backed securities, decrease the treasury bond yield, and widen the swap spread. Inversely, increases in interest rates can increase the price of MBS, increase the treasury bond yield, and narrow the swap spread. Such movements can be modeled through market participants response in buying/selling fixed legs of MBS swaps.

\section{Future Considerations}

In future studies, we wish to test the feasibility of the Vasicek model for term structure. In particular, we wish to determine the effectiveness in determining the change in swap spread due to negative duration hedging. We suspect that the model is insufficient for capturing all sources of market risk, and that macroeconomic, systemic and credit factors will be necessary. In modern practice, market participants use modified versions of the Heath-Jarrow-Morton (HJM) model or the multi-factor Hull-White model. There is also the unexplored possibility of using a modified version of the Cox-Ingersoll-Ross stochastic model for interest rates. To test the effectiveness of the model, 6-month, 12-month, 5-year, 15-year, and 30-year term structures will be used. Both the credit markets and mortgage markets are fascinating, and it remains to be seen the effects of efficient refinancing on the asset pricing of underlying instruments.

\end{document}